# Towards Impedance Characterization of Carbon-Carbon Ultrasonically Absorptive Coatings via the Inverse Helmholtz Problem

Danish Patel\*, Prateek Gupta\*, and Carlo Scalo†

*Purdue University, West Lafayette, Indiana 47907, USA*

Thomas Rothermel,‡ and Markus Kuhn‡

*German Aerospace Center (DLR), Stuttgart, Germany*

**We present a numerical method to determine the complex acoustic impedance at the open surface of an arbitrarily shaped cavity, associated to an impinging planar acoustic wave with a given wavenumber vector and frequency. We have achieved this by developing the first inverse Helmholtz Solver (iHS), which implicitly reconstructs the complex acoustic waveform–at a given frequency–up to the unknown impedance boundary, hereby providing the spatial distribution of impedance as a result of the calculation for that given frequency. We show that the algebraic closure conditions required by the inverse Helmholtz problem are physically related to the assignment of the spatial distribution of the pressure phase over the unknown impedance boundary. The iHS is embarassingly parallelizable over the frequency domain allowing for the straightforward determination of the full broadband impedance at every point of the target boundary. In this paper, we restrict our analysis to two-dimensions only. We first validate our results against Rott's quasi one-dimensional thermoacoustic theory for viscid and inviscid constant-area rectangular ducts, test our iHS in a fully unstructured fashion with a geometrically complex cavity, and finally, present a simplified, two-dimensional analysis of a sample of carbon-carbon ultrasonically absorptive coatings (C/C UACs) manufactured in DLR-Stuttgart, and used in the hypersonic transition delay experiments by Wagner et al. in AIAA 2012-5865. The final goal is to model C/C UACs with multi-oscillator Time Domain Impedance Boundary Conditions (TDIBC) developed by Lin et al. in JFM (2016) to be applied in direct numerical simulations (DNS) focused on the overlying boundary layer, eliminating the need to simultaneously resolve the microscopic porous structure of the C/C UACs.**

## I. Introduction

Boundary layers undergo laminar-to-turbulent transition as a result of the development and growth of perturbations due to a variety of factors including, but not limited to, free-stream turbulence, free-stream Mach number, surface-to-free-stream temperature ratio, surface roughness, surface curvature and sound radiation.[1,2] In the case of hypersonic boundary layers with small free-stream disturbances evolving over smooth surfaces and slender geometries, Mack modes govern the transition dynamics.[3,4] Among them, second-mode waves are of particular importance: they are manifest as convectively amplified acoustic modes resonating inside the boundary layer with frequencies typically in the ultrasonic range.[5] As a result, several passive control techniques for hypersonic boundary layer transition delay have been focusing on the design of ultrasonically absorptive coatings, or UAC(figure 1).[6] They have been shown to attenuate resonant modes[7] via thermo-viscous attenuation of waves trapped in the coatings' cavities, and, therefore, to delay transition.

---

\*Graduate Research Assistant, School of Mechanical Engineering
†Assistant Professor, School of Mechanical Engineering, scalo@purdue.edu
‡Research Scientist, Dep. Space System Integration

Acoustic energy absorption by UACs can be conveniently modeled by assigning equivalent impedance boundary conditions. In general, the (normal) acoustic impedance at a surface, $\mathcal{S}$, is defined as,

$$Z(\mathbf{x};\omega) = \frac{\hat{p}(\mathbf{x};\omega)}{\hat{\mathbf{U}}(\mathbf{x};\omega)\cdot\hat{\mathbf{n}}(\mathbf{x})} \quad \forall \quad x \in \mathcal{S}$$

where $\hat{\mathbf{U}}$ and $\hat{p}$ are Fourier transforms of the velocity and pressure at the surface, respectively, $\hat{\mathbf{n}}$ is the surface normal directed away from the flow side and into the cavity, $\mathbf{x}$ is the position vector on the surface, and $\omega$ is the angular frequency. Hereafter, vectorial quantities and/or arrays will be indicated in bold. For a quiescent fluid, the acoustic power transmitted through a surface—due to interaction with the constituent frequencies contained in an incident wave packet—with given impedance is,

$$\dot{W}_{ac} = \int_{-\infty}^{\infty} \left\{ \int_{\mathcal{S}} \frac{1}{2}\text{Re}\left[Z(\omega)\right] \left|\hat{\mathbf{U}}(\omega)\cdot\hat{\mathbf{n}}\right|^2 dA \right\} d\omega. \quad (1)$$

The acoustic impedance at a surface retains all the necessary information regarding the two-way acoustic coupling between waves on either side of the surface. As such, if the acoustic impedance were to be imposed, it would control the direction and magnitude of acoustic energy transfer.

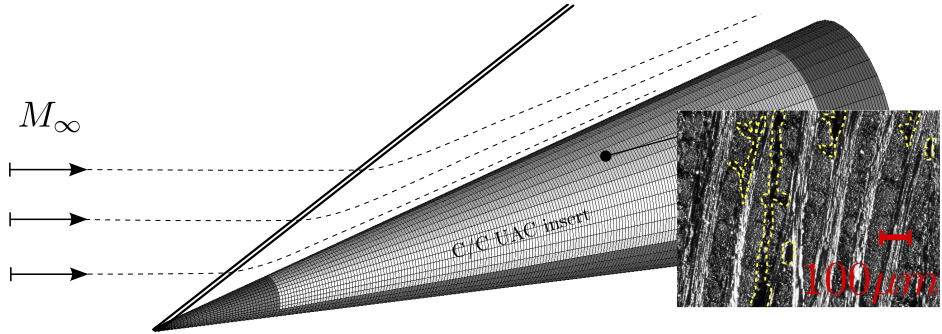

Figure 1: Illustration of a 7° half-angle round-tip cone model equipped with C/C UAC employed by Wagner[4] to damp second mode instabilities. The microporous structure of such coatings is highly irregular and only statistically predictable (see inset figure).

In this work, we propose a strategy to model open cavities such as the ones present in UACs by calculating their equivalent acoustic impedance (figure 2). The goal is then to fit appropriate multi-oscillator TDIBCs to the calculated impedance following the approach by Lin et al. (2016),[8] and apply them in fully compressible Navier-Stokes solvers focused only on resolving the overlying transitional boundary layer, hence saving significant computational resources at every time-step compared to pore-resolved cases.

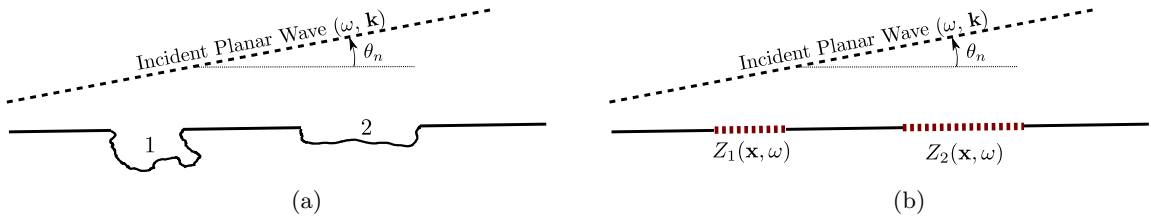

Figure 2: Illustration of a plane wave impinging on a porous surface (a) with closed cavities, which can be modeled by assigning equivalent impedance boundary conditions (IBCs) (b).

We note that this cannot be achieved with a classic acoustic analysis based on Helmholtz solvers. Such solvers rely on the formulation of an eigenvalue problem which yields a set of eigenmodes and corresponding complex eigenvalues containing frequency and growth/decay rate information. These solvers require conditions, and especially impedance, to be specified at *all boundaries* (figure 3a). We rather need a tool to analyze a cavity with a single open boundary where impedance is to be determined (figure 3b) and not specified or known a priori.

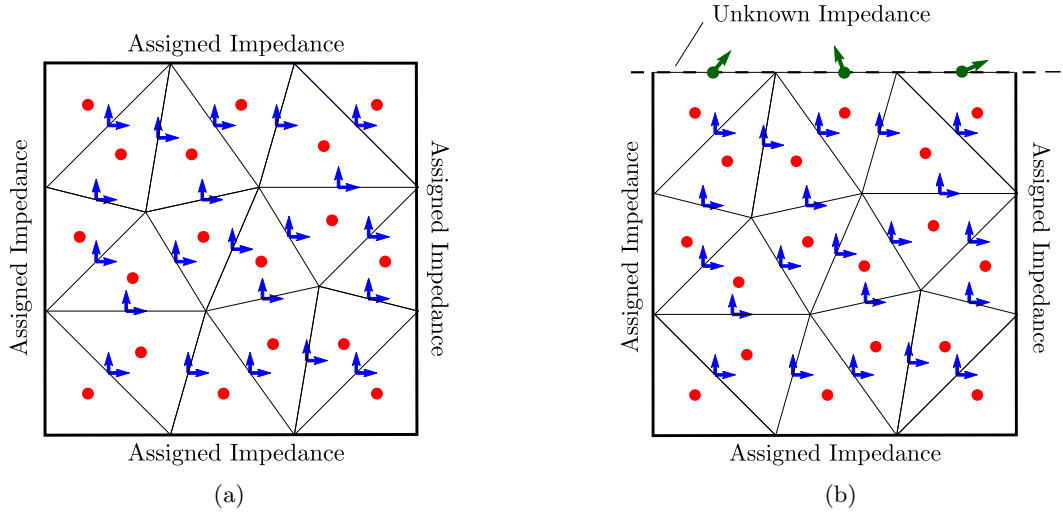

Figure 3: Sample two-dimensional computational setup for (a) direct and (b) inverse Helmholtz solvers. The direct solver requires impedance at all boundaries to be specified, whereas the inverse solver is able to provide the spatial distribution of the impedance at an *open* boundary as an outcome of the calculation.

In the current study, we achieve this by constructing the first inverse Helmholtz Solver (iHS). We begin with the linearized governing equations for a compressible fluid, coupled with a generalized equation of state, which are then cast into the inverse Helmholtz formulation in the frequency domain where part of the governing equations are extended to the unknown impedance boundary (section II). This problem is, however, rank deficient,[9] i.e. unclosed. This issue is resolved by applying appropriate—and physically realizable—algebraic closure conditions which are also discussed below. We then show validation of the iHS based on the acoustic impedance prediction for a constant-area rectangular duct against Rott's linear thermoacoustic wave theory for both viscid and inviscid test cases (section III). We then present the analysis of an idealized complex toy geometry (section IV), and finally, perform a simplified two-dimensional study of the acoustic absorption characteristics of a real C/C UAC geometry provided by DLR-Stuttgart applying the proposed inverse Helmholtz Solver methodology to a single cavity (section V).

## II.  Methodology: Inverse Helmholtz Solver (iHS)

In this section, we outline the inverse Helmholtz problem formulation starting with the linearization of the compressible flow governing equations for a fluid with a general equation of state. Although a more comprehensive derivation with mean flow and gradients is possible, we restrict the scope of this paper to a uniform, quiescent base flow and two-dimensional geometries without loss of generality for the iHS methodology.

### II.A.  Governing Equations

The fully non-linear governing equations for a general compressible fluid are,

$$\frac{\partial}{\partial t}(\rho) + \frac{\partial}{\partial x_j}(\rho u_j) = 0, \tag{2a}$$

$$\frac{\partial}{\partial t}(\rho u_i) + \frac{\partial}{\partial x_j}(\rho u_i u_j) = -\frac{\partial}{\partial x_i}p + \frac{\partial}{\partial x_j}\tau_{ij}, \tag{2b}$$

$$\frac{\partial}{\partial t}(\rho e) + \frac{\partial}{\partial x_j}[u_j(\rho e + p)] = \frac{\partial}{\partial x_j}(u_i \tau_{ij} - q_j), \tag{2c}$$

where $x_1$, $x_2$, and $x_3$ (or $x$, $y$ and $z$) are Cartesian coordinates, $u_i$ are the velocity components in each of these directions, and $p$, $\rho$, and $e$ are, respectively, the pressure, density, and internal energy per unit mass.

The shear stress $\tau_{ij}$, and heat flux $q_j$, are defined as,

$$\tau_{ij} = 2\mu \frac{\partial u_j}{\partial x_i} + \lambda \frac{\partial u_k}{\partial x_k} \delta_{ij},$$

$$q_j = -\kappa \frac{\partial T}{\partial x_j},$$

where $\mu$ and $\lambda$ are the first and second viscosity coefficients respectively and $\kappa$ is the coefficient of thermal conductivity. The field variables are decomposed as,

$$\begin{aligned} p(\mathbf{x};t) &= p_0 + p'(\mathbf{x};t) \\ T(\mathbf{x};t) &= T_0 + T'(\mathbf{x};t) \\ \rho(\mathbf{x};t) &= \rho_0 + \rho'(\mathbf{x};t) \\ e(\mathbf{x};t) &= e_0 + e'(\mathbf{x};t) \\ u_i(\mathbf{x};t) &= u'_i(\mathbf{x};t) \end{aligned} \qquad (3)$$

where the subscript 0 denotes the base (quiescent) state and the primes denote fluctuations. Upon substitution of the equations (3) into the system of equations (2) and dropping terms higher than first order in primed variables, we obtain the linearized equations,

$$\frac{\partial \rho'}{\partial t} + \rho_0 \frac{\partial u'_k}{\partial x_k} = 0, \qquad (4a)$$

$$\rho_0 \frac{\partial e'}{\partial t} + e_0 \frac{\partial \rho'}{\partial t} + \rho_0 e_0 \frac{\partial u'_k}{\partial x_k} + p_0 \frac{\partial u'_k}{\partial x_k} - \kappa \frac{\partial}{\partial x_k}\left(\frac{\partial}{\partial x_k} T'\right) = 0, \qquad (4b)$$

$$\rho_0 \frac{\partial u'_i}{\partial t} + \frac{\partial p'}{\partial x_i} = \frac{\partial}{\partial x_j} \tau'_{ij}. \qquad (4c)$$

In the general theory for linear acoustics, primary thermoviscous losses are captured via terms containing the dynamic viscosity, $\mu$ in $\tau_{ij}$ and the thermal conductivity, $\kappa$. In addition to these losses, in the ultrasonic regime, bulk viscous losses become significant and are accounted for in $\lambda$, the coefficient of second viscosity, which can be calculated from the bulk viscosity, $\mu_b \equiv \lambda + (2/3)\mu$. For the latter, we refer to the findings by Lin et al.[10] where the bulk viscosity, and hence the second viscosity, $\lambda$, can be expressed as a function of the frequency, base pressure and temperature.

The total differentials of pressure and internal energy for a general equation of state read,

$$dp = \frac{1}{\rho \beta_T} d\rho + \frac{\alpha_v}{\beta_T} dT \qquad (5)$$

$$de = c_v dT - B_0 d\rho \qquad (6)$$

with, $\quad \alpha_v = -\dfrac{\left(\rho \frac{\partial p}{\partial T}\big|_v\right)}{\left(\frac{\partial p}{\partial v}\big|_T\right)}, \quad \beta_T = -\left(v \frac{\partial p}{\partial v}\big|_T\right)^{-1}, \quad B_0 = \dfrac{C_0}{(\rho_0)^2} \quad$ and, $\quad C_0 = \left(T_0 \frac{\partial p_0}{\partial T_0}\big|_v - p_0\right),$

where the coefficient of thermal expansion, $\alpha_v$, the isothermal compressibility, $\beta_T$, and constants $B_0$ and $C_0$ can be obtained analytically from any given equation of state. For a calorically perfect ideal gas, these variables revert to,

$$\alpha_v = \frac{1}{T_0} \qquad \beta_T = \frac{1}{p_0} \qquad B_0 = C_0 = 0.$$

Substituting (5) and (6) in equation (4) and eliminating density perturbations yields,

$$\rho_0 \beta_T \frac{\partial p'}{\partial t} - \rho_0 \alpha_v \frac{\partial T'}{\partial t} + \rho_0 \frac{\partial u'_k}{\partial x_k} = 0, \qquad (7a)$$

$$(\rho_0 c_v + C_0 \alpha_v) \frac{\partial T'}{\partial t} - C_0 \beta_T \frac{\partial p'}{\partial t} + p_0 \frac{\partial u'_k}{\partial x_k} - \kappa \frac{\partial}{\partial x_k}\left(\frac{\partial}{\partial x_k} T'\right) = 0, \qquad (7b)$$

$$\rho_0 \frac{\partial u'_i}{\partial t} + \frac{\partial p'}{\partial x_i} = \mu \frac{\partial}{\partial x_k}\left(\frac{\partial}{\partial x_k} u'_i\right) + (\mu + \lambda) \frac{\partial}{\partial x_i}\left(\frac{\partial u'_k}{\partial x_k}\right), \qquad (7c)$$

which is the final form that will be discretized.

## II.B. Numerics and Algebraic Formulation

Converting the system of equations (7) from time domain to frequency domain, collecting all terms containing the frequency, $\omega$, and applying a spatial discretization yields the following system of linear equations,

$$i\omega \mathbf{X_0} + \mathbf{A_{eig}} \mathbf{X_0} = \mathbf{b} \tag{8}$$

where,

$$\mathbf{X_0} = \begin{pmatrix} \hat{\mathbf{p}} & \hat{\mathbf{T}} & \hat{\mathbf{u}}_1 & \hat{\mathbf{u}}_2 & \hat{\mathbf{u}}_3 \end{pmatrix}^{\mathbf{T}} \quad \text{and,} \quad \mathbf{b} = \mathbf{0}.$$

We adopt a second order discretization wise staggered variable arrangement on an unstructured grid (shown in figure 4), where thermodynamic field variables (pressure $p$, specific internal energy $e$, and hence temperature $T$ and density $\rho$) are defined at the cell centers and velocity field components $u_i$ at the face centers.

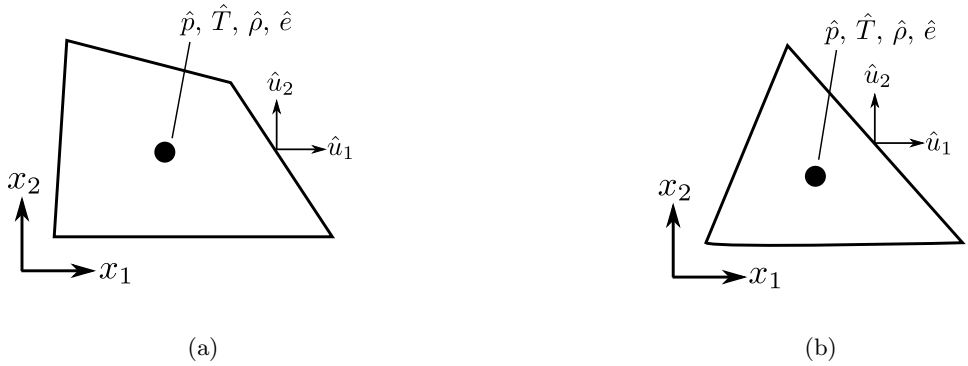

Figure 4: Staggered variable arrangement over a (a) QUAD, and (b) TRI cell. All thermodynamic variables, $p$, $T$, $\rho$, $e$ and $c_v$, are collocated on cell centers ($\bullet$), while velocity components, $u_i$, are located on face centers

Equation (8) with appropriate boundary conditions would yield a conventional Helmholtz problem, which is essentially an eigenvalue problem that is typically considered in linear acoustic analysis.[11] Solving this problem returns a finite set of frequencies (eigenvalues) and wave forms (eigenvectors). Note that homogeneous boundary conditions ($\mathbf{b} = \mathbf{0}$) are required by an eigenvalue solver.

On the other hand, in the inverse Helmholtz solver approach, the vector $\mathbf{b}$ in equation (8) is no longer null as it contains closure conditions (explained below) to be applied at the unknown impedance boundary (termed **IBC**). The value of $\omega$ is now real and a given, and the system (9) is solved for an extended set of unknowns, which include velocities and pressures at the boundary ($\hat{\mathbf{u}}_{\mathbf{1,IBC}}$, $\hat{\mathbf{u}}_{\mathbf{2,IBC}}$, $\hat{\mathbf{u}}_{\mathbf{3,IBC}}$ and $\hat{\mathbf{p}}_{\mathbf{IBC}}$), in addition to the previous unknowns in $\mathbf{X_0}$. The momentum equation is now written at the **IBC** and is symbolically represented by blocks **I** and **II** in the overall inverse Helmholtz problem formulation, which reads,

$$i\omega \begin{pmatrix} \hat{\mathbf{p}} \\ \hat{\mathbf{T}} \\ \hat{\mathbf{u}}_1 \\ \hat{\mathbf{u}}_2 \\ \hat{\mathbf{u}}_3 \\ \hat{\mathbf{u}}_{1,IBC} \\ \hat{\mathbf{u}}_{2,IBC} \\ \hat{\mathbf{u}}_{3,IBC} \\ \hat{\mathbf{p}}_{IBC} \end{pmatrix} + \begin{pmatrix} \boxed{\mathbf{A_{eig}} \quad \mathbf{II}} \\ \boxed{\mathbf{I}} \\ \boxed{\mathbf{III}} \end{pmatrix} \begin{pmatrix} \hat{\mathbf{p}} \\ \hat{\mathbf{T}} \\ \hat{\mathbf{u}}_1 \\ \hat{\mathbf{u}}_2 \\ \hat{\mathbf{u}}_3 \\ \hat{\mathbf{u}}_{1,IBC} \\ \hat{\mathbf{u}}_{2,IBC} \\ \hat{\mathbf{u}}_{3,IBC} \\ \hat{\mathbf{p}}_{IBC} \end{pmatrix} = \begin{pmatrix} \mathbf{b_p} \\ \mathbf{b_T} \\ \vdots \\ \vdots \\ \vdots \\ \vdots \\ \vdots \\ \vdots \\ \mathbf{b_{IBC}} \end{pmatrix}. \tag{9}$$

In order to close the system, we introduce a pressure variable, $\hat{\mathbf{p}}_{\mathbf{IBC}}$ at the **IBC**. This later allows us to evaluate directly the value of impedance at the **IBC**, without having to rely on extrapolation techniques that may be inaccurate for complex geometries.

Closure conditions are formulated as,

$$\frac{\hat{\mathbf{p}}_{\mathbf{IBC,m+1}}}{\hat{\mathbf{p}}_{\mathbf{IBC,m}}} = \Psi_m, \tag{10}$$

symbolically represented by block **III** of the system (9), and can be physically related to the shape of the impinging wave. For example, in the case of a two-dimensional planar wave of the shape, $p_{\mathbf{wave}} = p_\infty \exp\left(-ik\left(x\cos(\theta_n) + y\sin(\theta_n)\right)\right)$—where, $p_\infty$ is the amplitude, $k$ the wave number and, $\theta_n$ the angle of incidence of the incident wave on the unknown impedance boundary—equation (10) becomes,

$$\frac{\hat{\mathbf{p}}_{\mathbf{IBC,m+1}}}{\hat{\mathbf{p}}_{\mathbf{IBC,m}}} = \exp\left(-i\left\{k\cos(\theta_n)\left(x_{\mathbf{IBC,m+1}} - x_{\mathbf{IBC,m}}\right) + k\sin(\theta_n)\left(y_{\mathbf{IBC,m+1}} - y_{\mathbf{IBC,m}}\right)\right\}\right) = \Psi_m \tag{11}$$

Finally, to achieve full rank in the system we assign an arbitrary reference pressure, $\hat{p}_{\text{ref}}$ in an arbitrarily chosen cell corresponding to one row of block **III**. We note that the final value of impedance is independent from the specific value of the reference pressure since the former is a ratio of two quantities, both proportional to $\hat{p}_{\text{ref}}$.

These two closure conditions are finally incorporated into the system of equations (9), restoring its full rank, allowing to solve for the discrete pressure and velocity components at the **IBC** as well as inside the domain. Once we have obtained the complete solution, the impedance can directly be calculated as,

$$Z(\mathbf{x}; \omega) = \frac{\hat{\mathbf{p}}_{\mathbf{IBC}}}{\vec{\mathbf{u}}_{\mathbf{IBC}} \cdot \hat{\mathbf{n}}_{\mathbf{IBC}}} \tag{12}$$

where, $\vec{\mathbf{u}}_{\mathbf{IBC}} = (\hat{\mathbf{u}}_{\mathbf{1,IBC}}, \hat{\mathbf{u}}_{\mathbf{2,IBC}}, \hat{\mathbf{u}}_{\mathbf{3,IBC}})$ is the velocity vector and $\hat{\mathbf{n}}_{\mathbf{IBC}}$ the surface normal facing inside, towards the domain.

## III. Validation Against Rott's Thermoacoustic Theory

The basic working principles of the inverse Helmholtz Solver (iHS), introduced in a general form in the previous section, are here explained with a simple 1-D numerical example.

Using a staggered grid approach with a set frequency, $\omega$, Rott's wave equations,[12,13]

$$i\omega \hat{p} = -\frac{1}{1 + (\gamma - 1)f_\kappa} \rho_0 a_0^2 \frac{d\hat{u}}{dx} \tag{13a}$$

$$i\omega \hat{U} = -(1 - f_\nu) A \frac{1}{\rho_0} \frac{d\hat{p}}{dx} \tag{13b}$$

with,

$$f_\nu = \frac{\tanh\left(\frac{(1+i)y_0}{\sqrt{2\nu/\omega}}\right)}{\left(\frac{(1+i)y_0}{\sqrt{2\nu/\omega}}\right)}, \qquad f_\kappa = \frac{\tanh\left(\frac{(1+i)y_0}{\delta_\nu/\sqrt{Pr}}\right)}{\left(\frac{(1+i)y_0}{\delta_\nu/\sqrt{Pr}}\right)}, \qquad \text{and} \qquad y_0 = \frac{h}{2}$$

can be spatially integrated from one hard end (on the left in figure 5), starting with a given arbitrary pressure value, all the way to the unknown impedance boundary at $x = H$. Here, $\hat{p}$ represents discrete pressure perturbations and $\hat{U}$ represents the volume flow rate through a constant cross-sectional area $A = h \times 1$, where unit depth in $z$ axis has been assumed. These equations account for viscous losses at the wall (via the thermoviscous functions $f_\kappa$ and $f_\nu$) and represent the conservation of mass and energy combined (13a), and momentum (13b). Following this procedure, the impedance at the end of the duct for any given frequency is a *result* of the calculation and not an input as it would be for a classic acoustic eigenvalue problem.

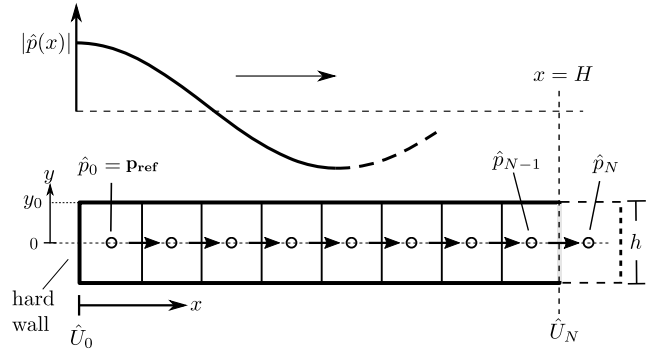

Figure 5: Inverse wave reconstruction via one dimensional spatial integration of Rott's wave equations. The impedance at the end of the tube is calculated directly via $Z(x;\omega) = \frac{1}{2}(\hat{p}_{N-1} + \hat{p}_N)A/\hat{U}_N$.

The cross-sectional averaged velocity at the end of the duct, $\hat{U}(x = H) = \hat{U}_N$ can be used to recover the full velocity profile in the wall-normal direction using the equation:

$$\hat{u}(x = H, y) = \frac{\hat{U}_N}{A}\left(1 - \frac{\cosh(\eta)}{\cosh(\eta_0)}\right) \tag{14}$$

$$\eta_0 = y_0\sqrt{\frac{i\omega}{\nu}}, \quad \text{and} \quad \eta(y) = y\sqrt{\frac{i\omega}{\nu}} \quad \text{with} \quad 0 \leq y < y_0.$$

The wall-normal profile of acoustic impedance at the end of the duct for each frequency, $\omega$, can then be calculated as,

$$Z(\mathbf{x}, \omega) = \frac{\frac{1}{2}(\hat{p}_{N-1} + \hat{p}_N)}{\hat{u}(x = H, y)}. \tag{15}$$

Note that in the absence of viscosity, $f_\nu$ and $f_\kappa$ should be explicitly set to zero and the equations (13) revert to those for isentropic acoustics.

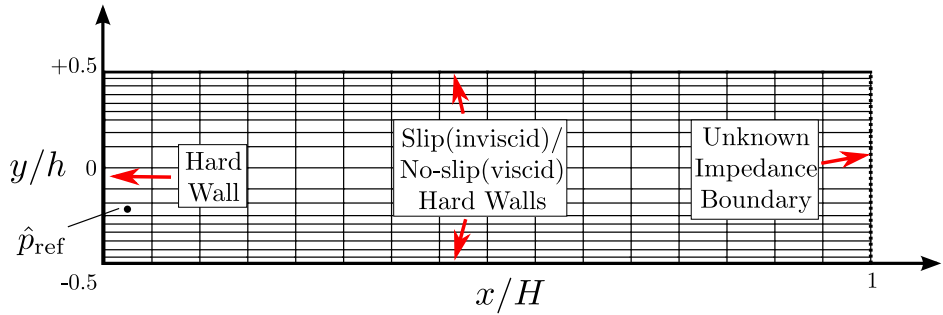

Figure 6: Computational set up to analyze a simple duct using the inverse Helmholtz Solver (iHS). The grid shown here is just meant for illustrative purposes.

Figure 6 shows the two-dimensional computational set up used in the iHS runs, compared against Rott's theory in the following subsections. The dimensions of the duct are $H = 4$ mm and $h = 1$ mm. Results are first obtained for an inviscid duct, with lateral slip walls for a range of frequencies in the ultrasonic regime, and then for a viscous duct with a viscosity of $\nu = 0.05$ m$^2$/s.

### III.A.   Inviscid Duct

Figure 7 compares solutions obtained from the iHS against semi-analytical solutions obtained from Rott's wave equations over several decades of frequencies for an inviscid duct. In the absence of shear stress, no wave attenuation should be present (if not numerical) and the acoustic velocity profile is flat everywhere. In figure 7, the periodic admittance peaks represent resonant frequencies at which the duct length is a multiple integer of the acoustic wavelength, and where the surface normal admittance at the end of the duct diverges.

The quality of the agreement decreases at higher frequencies where the numerical solution eventually becomes under-resolved in the axial direction.

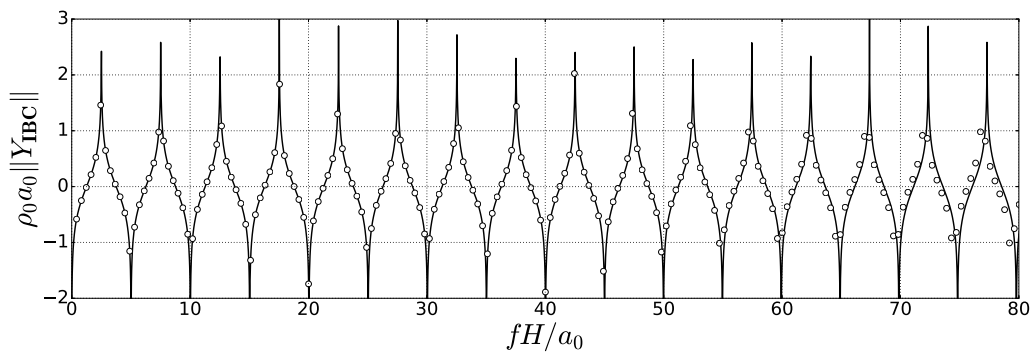

Figure 7: Admittance magnitude obtained from the iHS (—) and from Rott's wave equations (○) for an inviscid two-dimensional duct.

### III.B. Viscous Duct

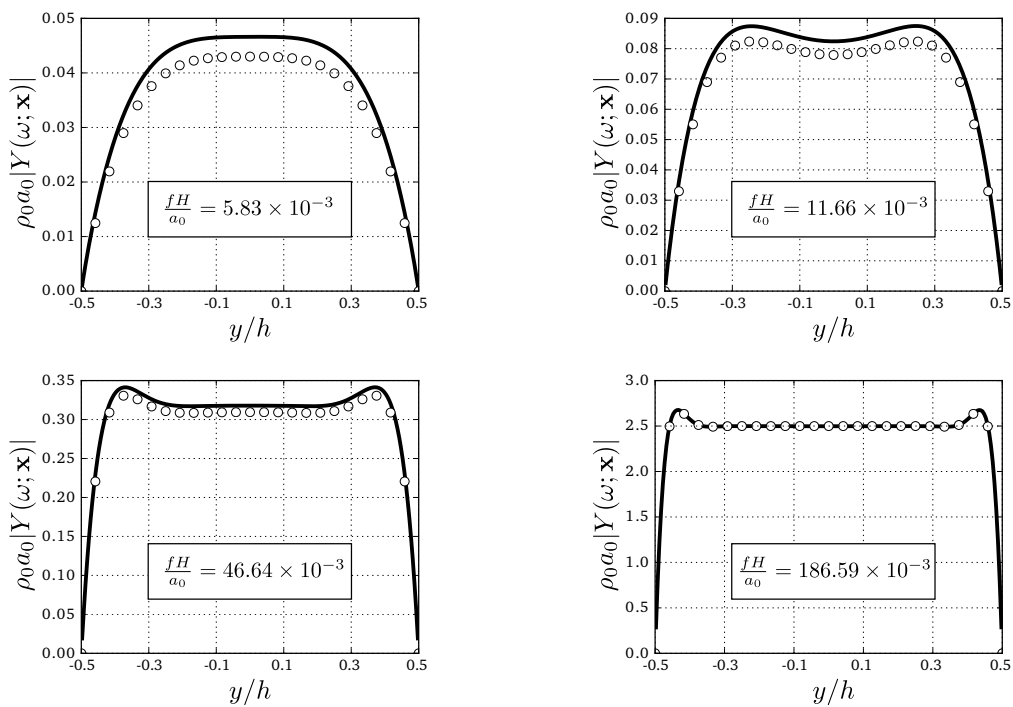

Figure 8: Comparisons of results obtained from the iHS (—) and Rott's wave equations (○) for a two-dimensional viscid duct at different frequencies.

Figure 8 shows how solutions from the iHS compare with Rott's thermoacoustic theory for viscous cases. It is observed that at lower frequencies there is poor matching with the semi-analytical solution, primarily due to the fact that Rott's thermoacoustic theory chooses to treat the duct aspect ratio as very large, not accounting for edge effects, whereas in our case we have $H/h = 4$. At high frequencies, edge effects are confined near the hard end and away from the unknown impedance boundary, returning perfect matching with Rott's solutions. In general, it was found that there is good agreement with Rott's solutions at frequencies where the Stokes boundary layer thickness was less than 20 % of the duct width for this specific geometry.

## IV. Two-Dimensional Toy Cavity

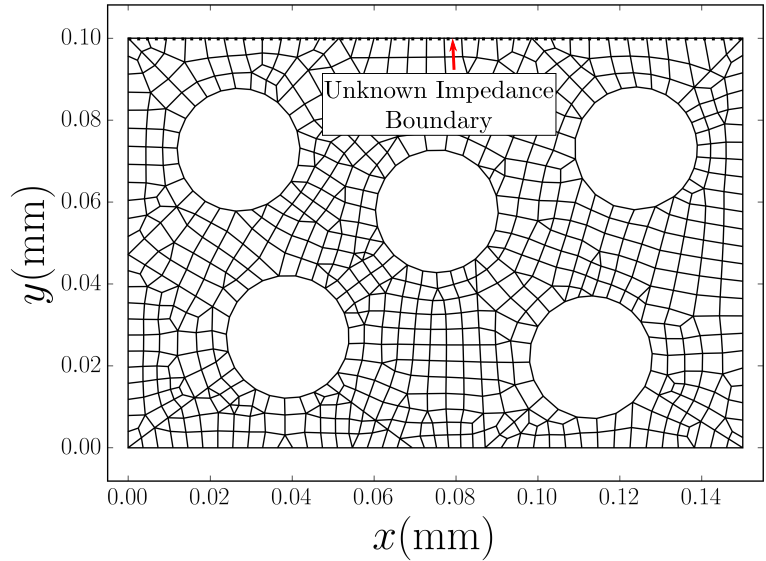

Figure 9: Sample unstructured computational mesh discretizing idealized toy cavity geometry analyzed using the iHS.

In this section, we present the analysis of an idealized UAC geometry (figure 9) using the inverse Helmholtz solver developed in the previous sections. The impinging wave is taken to be a planar wave that is first oriented normally (figure 10a), and then at an angle of $135^o$ measured with respect to the positive $x$ axis (figure 10b).

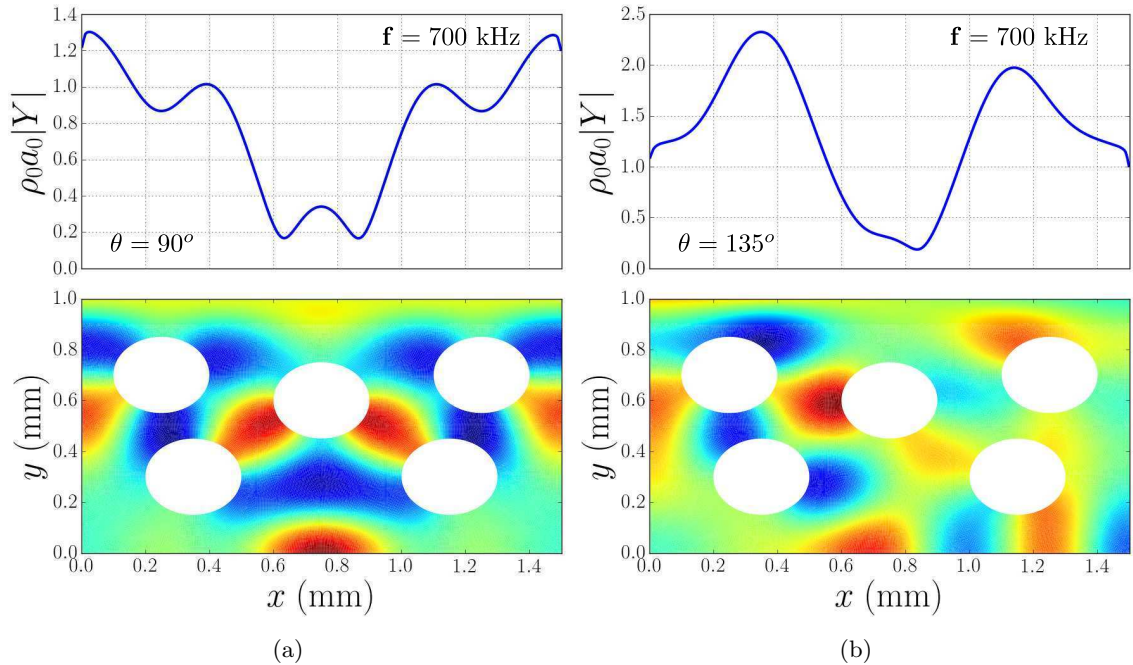

Figure 10: Pressure waveform and magnitude of normal admittance for a 700 kHz planar wave for $\theta = 90^o$ (a) and $\theta = 135^o$ (b), where $\theta$ is the wave's incidence angle measured with respect to the positive $x$ axis.

Figure 10 shows the pressure contours and surface normal specific admittance magnitude at the impedance boundary of the idealized geometry, in response to a planar wave. In both cases, the viscous boundary layers responsible for wave attenuation are resolved. While the admittance at the edges–where the **IBC** intersects with the hard, no-slip wall–is numerically almost the same for both cases, the intermediate profiles are different. In the normal case, the admittance is lowest near the center obstacle, due to interference from waves reflected off the obstacles on either side of it. For the angled case, the minima occurs approximately in the region bounded by the three obstacles on the right. The peak admittance in the angled case is significantly higher than that in the case of normal incidence, due to the fact that the angled wave is able to traverse more distance before it encounters the obstacles on the right.

## V. Two-Dimensional Analysis of C/C UAC

In this section, we employ the iHS to analyze the impedance of C/C UACs.

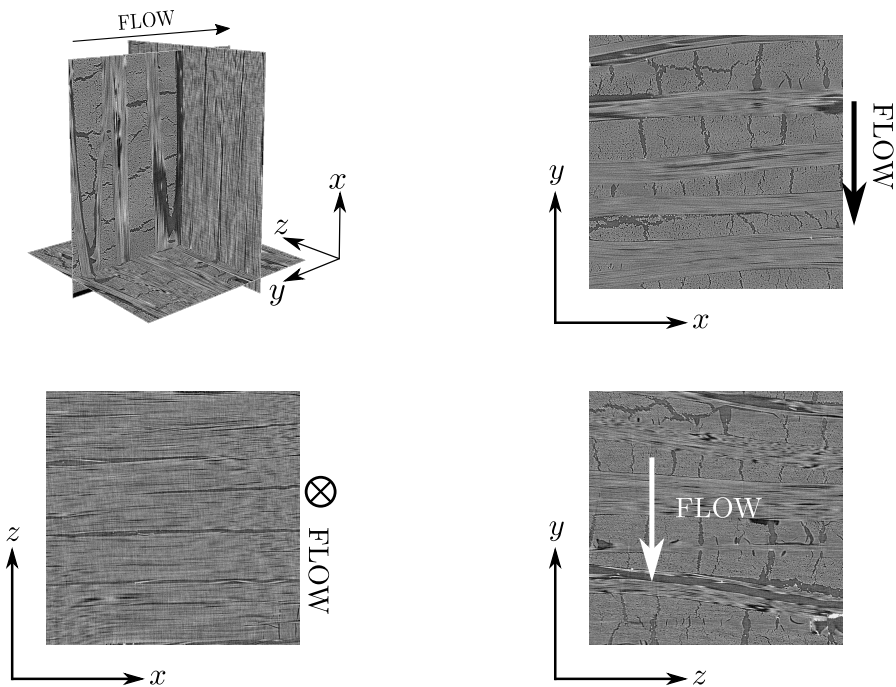

Figure 11: Two-dimensional slices of a subvolume of the three dimensional C/C UAC insert shown in figure 1, each 1.5 mm × 1.5 mm in size.

Carbon fiber reinforced carbon ceramic (C/C) belongs to a class of materials commonly known as ceramic matrix composites (CMCs). C/C CMCs have excellent thermal resistance and a low coefficient of thermal expansion, rendering them ideal for structural uses at high temperatures typically encountered in re-entry vehicles.[14,15] Moreover, porous C/C surfaces have been shown to damp second-mode instabilities in hypersonic boundary layers, delaying the laminar-to-turbulent trasition.[16–18]

C/C originates as a semifinished product from the C/C-SiC manufacturing process via the liquid silicone infiltration route as described by Krenkel.[19] The entire process is depicted in figure 13. The manufacturing comprises the fabrication of a green body of carbon fiber reinforced plastic (CFRP), in this case consisting of carbon fibers impregnated with a phenolic resin processed via autoclave technique. The green body is cured and subsequently pyrolyzed at temperatures up to 1650 C. This process step converts the matrix component from phenolic resin to amorphous carbon. During the following cool-down the shrinkage of the pyrolyzed matrix is hindered by the carbon fibers resulting in a pattern of micro cracks.

The micro crack system of the C/C CMC used here is shown in figure 11. The tomographic images were taken with the aid of a GE Phoenix Nanotom© 180nF computer tomograph, using a voxel size of 1 $\mu$m and a volume of interest (VOI) of $1500^3$ voxels.

It can be seen that the C/C is based on a CRFP with a 0—90 lay-up of carbon fiber fabric. Therefore, the material exhibits an orthotropic layout and behaviour, resulting in different thermo-mechanical and flow-through properties in plane (x, z) and transverse (y) w.r.t. the lay-up.

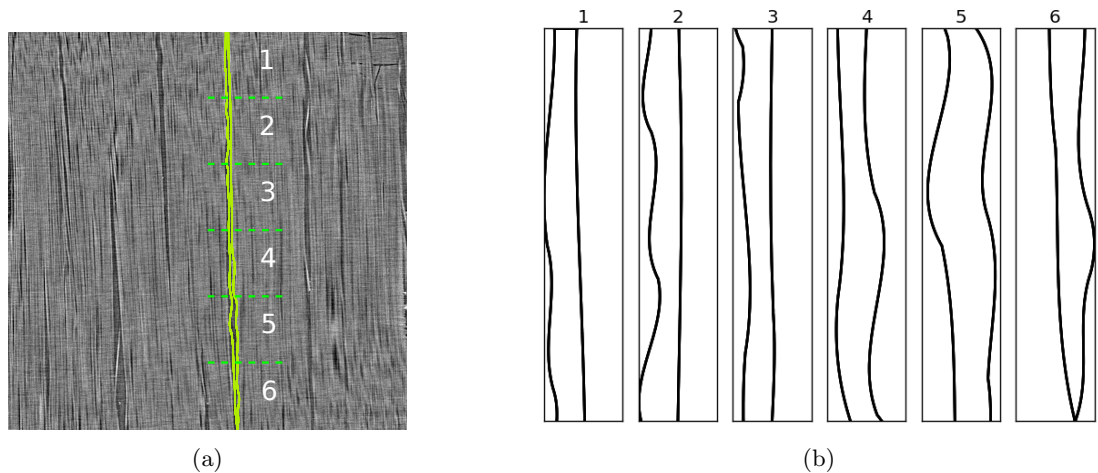

(a)          (b)

Figure 12: Single cavity extracted from the X-Z plane (a), reconstructed edge of analyzed cavity segmented into 6 parts (b).

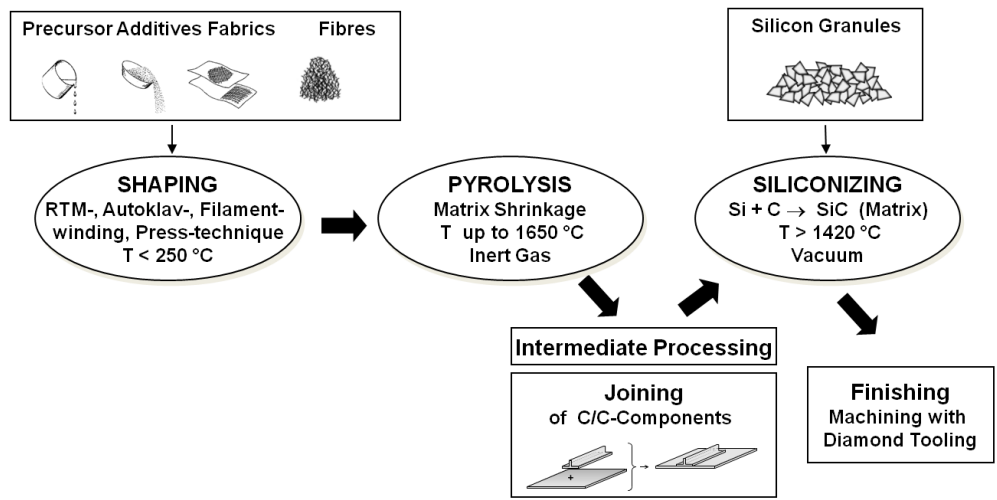

Figure 13: Manufacturing of C/C-SiC via LSI-process

Figure 11 shows 2-D slices of a three dimensional C/C block comprising numerous microporous cavities (as shown in figure 1). We extracted one cavity from the X-Z plane (figure 12a) and studied the variation of acoustic admitttance while varying the frequency in the ultrasonic range for a normally incident plane wave. Figure 12b shows six constituent slices of the extracted cavity.

Figure 14 shows the acoustic pressure response of the cavity to a normally incident planar wave at various frequencies. As part of the required closure conditions, we assign a reference pressure $\hat{p}_{\text{ref}} = 1 + 1j$ to a cell at the lower end of the cavity (figure 12b), which is also used to scale the pressure waveforms. In all cases, a standing wave extending depth-wise in the cavity is observed, decaying in amplitude due to viscous attenuation. In general, penetration depth decreases with increasing frequency, while attenuation rate increases. Figure 15 shows the variation of the admittance profile over the mouth of the cavity for frequencies, $f = 100, 300, 600$ and $1000$ kHz. The slender cavity exhibits wave profile characteristics similar to those of a rectangular duct with lateral size comparable to Stokes boundary layer thickness, resulting in a parabolic profile for admittance. In general, the admittance magnitude is monotonically increasing with frequency.

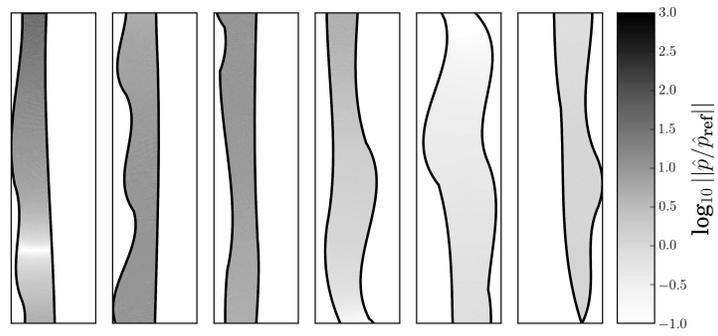

(a) 100 kHz

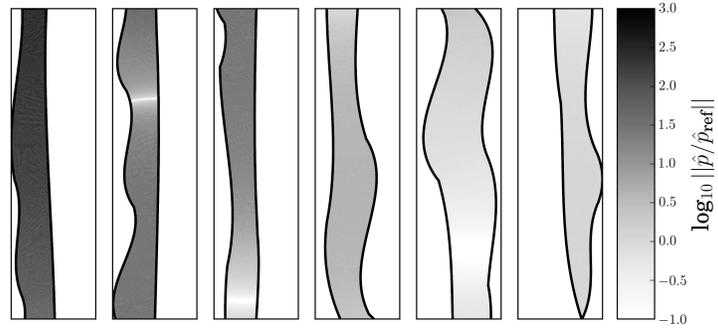

(b) 300 kHz

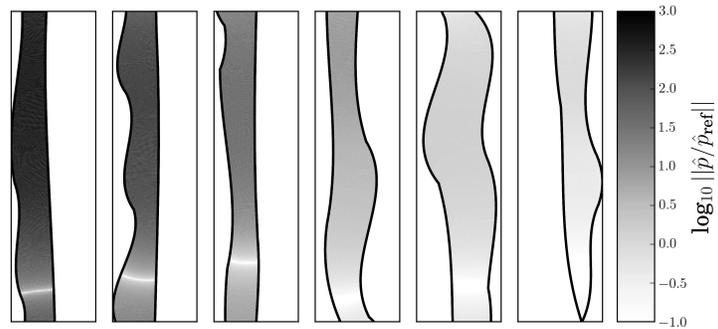

(c) 600 kHz

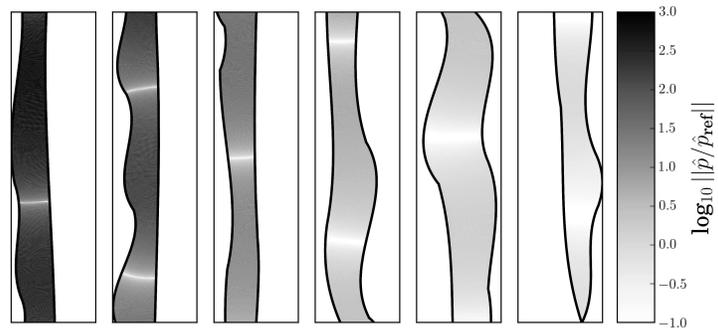

(d) 1 MHz

Figure 14: Pressure waveforms induced by a normal incident plane wave for various frequencies in the range 100 kHz—1 MHz.

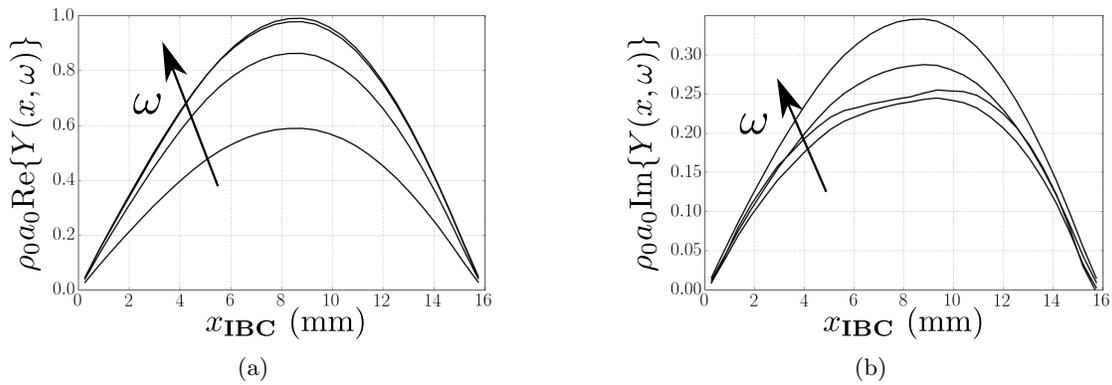

Figure 15: Profiles of (a) Real, and (b) Imaginary parts of the surface admittance for frequencies 100, 300, 600 and 1000 kHz.

## VI. Conclusion

We have presented an inverse Helmholtz Solver (iHS) methodology to determine the linear acoustic impedance at one or multiple open boundaries of an arbitrarily shaped cavity for an impinging planar wave at a given frequency. Utilizing the iHS methodology, we calculated the admittance for a two-dimensional open end duct for inviscid as well as viscid cases. This allowed us to provide direct comparisons with Rott's quasi one-dimensional thermoacoustic theory. In general, there is excellent matching with solutions obtained from Rott's wave equations in the absence of viscosity, and at high frequencies in the presence of viscosity owing to the diminishing effect of edge vortices. We then proceeded to analyze a toy ultrasonically absorptive cavity showcasing the solver's ability to capture significant non-uniformity in the admittance profiles for normal and angled incidence. Finally, we presented a simplified two-dimensional analysis of a real carbon-carbon ultrasonically absorptive coating. We found that the impinging waves penetrate deeper into the cavities at lower frequencies while decaying faster at higher frequencies.

Ongoing work is focused on extending the iHS analysis to geometrically complex three-dimensional cavities, with companion Navier-Stokes simulations based on multi oscillator Time Domain Impedance Boundary Conditions (TDIBCs). The developed iHS will be used to inform the design of a new generation of ultrasonically absorptive coatings.

## Acknowledgements

We acknowledge the support of the Rosen Center for Advanced Computing (RCAC) at Purdue, the Air Force Office of Scientific Research (AFOSR) grant FA9550-16-1-0209 and the very fruitful discussions with "Pon" R. Ponnappan (AFOSR), Dr. Ivett Leyva (AFOSR), Dr. Alex Wagner (DLR) and Dr. Viola Wartemann (DLR).

Additionally, the author Thomas Rothermel would like to thank the German Research Foundation (DFG) for financial support of the project within the Cluster of Excellence in Simulation Technology (EXC 310/1) at the University of Stuttgart.